\documentclass[final,5p,times,twocolumn]{elsarticle}
\biboptions{sort&compress}
\usepackage{lineno,isotope,mathrsfs}
\modulolinenumbers[5]
\usepackage{amsmath,amssymb,mathptmx,epsfig,bm,mathrsfs,feynmp,cuted,aas_macros}
\usepackage{color,slashed,exscale,multirow,soul,appendix}
\usepackage{times,txfonts,xcolor,graphicx,gensymb,tikz,tablefootnote,placeins}
\usepackage[normalem]{ulem}
\usepackage[breaklinks,colorlinks,citecolor=blue]{hyperref}
\newcommand{\bea}{\begin{eqnarray}}
\newcommand{\eea}{\end{eqnarray}}

\graphicspath{ {./figures/} } 

\newif\ifshowrevisions
\showrevisionstrue

\begin{document}
\begin{frontmatter}

\title{Covariant density functional from relativistic Brueckner theory with Chebyshev-expanded couplings}

\author[a,b]{Armen Sedrakian}
\ead{sedrakian@fias.uni-frankfurt.de}

\address[a]{Institute of Theoretical Physics, University of Wroc\l{}aw,
             50-204 Wroc\l{}aw, Poland}
\address[b]{Frankfurt Institute for Advanced Studies, 
             D-60438 Frankfurt am Main, Germany}
\begin{abstract}

We construct a density-dependent covariant energy density functional
(CE-ME1) from full-Dirac-space relativistic Brueckner--Hartree--Fock
(RBHF) self-energies and the corresponding equation of state. The
space-like component of the vector self-energy is accounted for
at the Fermi surface through reduced scalar and time-like
self-energies, which are mapped onto local
Hartree $\sigma$, $\omega$, $\delta$, and $\rho$ channels. The
density-dependent couplings are represented by cubic Chebyshev
polynomials over the density range $0.08\leq\rho\leq0.48~{\rm fm}^{-3}$. A constrained
energy-and-saturation calibration uses 199 RBHF energy-per-particle values
from published microscopic computations  at six isospin asymmetries, while
maintaining positive, monotonic couplings. The Bonn-A RBHF saturation
point, incompressibility $K_\infty=258$ MeV, symmetry energy $J=33.1$
MeV, and slope $L=65.2$ MeV are reproduced. The predicted curvature
$K_{\rm sym}=33$ MeV differs from the published RBHF value of $-67$
MeV, illustrating the derivative sensitivity of the local reduction.
The combined energy RMS is $0.613$ MeV, measuring model discrepancy
rather than statistical uncertainty. Uniform matter remains
mechanically stable and causal above saturation within the calibrated
range. In cold charge-neutral $npe\mu$ matter, CE-ME1 reasonably
reproduces the interpolated RBHF pressure and proton fraction over
most of the interval, although the pressure discrepancy increases
near its upper edge. The DD-ME2 benchmark is stiffer and predicts
a lower proton fraction at high density.
\end{abstract}


\end{frontmatter}

\section{Introduction}
\label{sec:introduction}

Covariant density functional (CDF) theory provides a successful
relativistic framework for finite nuclei, uniform nuclear matter,
and neutron-star matter~\cite{Serot1997,Ring1996,Meng2006,Sedrakian2023}.  In
density-dependent CDFs, many-body correlations are encoded in
effective meson--nucleon vertices that depend on the baryon density.
Microscopic relativistic Brueckner--Hartree--Fock (RBHF)
calculations provide guidance for this density dependence.
A particularly relevant precedent is the work of Typel and
Wolter~\cite{Typel1999}, who used
density-dependent $\sigma$, $\omega$, and $\rho$ vertices for
nuclear matter and finite nuclei and compared their behavior with
couplings inferred from Dirac--Brueckner self-energies. Subsequent
work further developed the microscopic-to-local connection
\cite{Dalen2005,Gogelein2008,Giai2010,Typel2018}. Conventional exponential
and rational parametrizations provide economical alternatives for
organizing the density dependence.

The purpose of the present work is to construct a new covariant
density functional, CE-ME1, through the reduction of modern
full-Dirac-space RBHF information to local Hartree fields. The
availability of all three Lorentz components $U_S$, $U_0$, and $U_V$
permits us to quantify the role of the spatial-vector self-energy in
this reduction, to construct meson-resolved density-dependent Hartree
couplings, and to test simultaneously their consistency with the
microscopic RBHF equation of state (EoS). For a compact representation
of the resulting density dependence of the vertices $g_i(\rho)$, we
use the Chebyshev expansion
\begin{equation}
C_i(\rho)\equiv\left(\frac{g_i(\rho)}{m_i}\right)^2
=\sum_{n=0}^{N_i}a_{in}T_n[z(\rho)],
\label{eq:cheb_intro}
\end{equation}
where $T_n$ are Chebyshev polynomials of the first kind,
$i=\sigma,\omega,\delta,\rho$ labels the meson channel, and $m_i$ is
its meson mass. In uniform matter the Hartree approximation implies
that only the combinations $C_i$, not the individual $g_i$ and $m_i$,
appear.  Chebyshev polynomials are chosen because their amplitudes
are uniformly bounded, $|T_n(z)|\leq1$ for $|z|\leq1$, and their
recurrence provides a numerically stable finite-interval representation
with analytic density derivatives. The low-order coefficients provide
smooth density dependence,  but we do not claim a mathematical advantage
over other orthogonal polynomial bases. A cubic Chebyshev expansion
spans precisely the same function space as any other cubic polynomial
basis; orthogonality under the continuous Chebyshev weight does not
make the nonlinear calibration coefficients independent.

The microscopic self-energies and energies used here were digitized
from previously published full-Dirac-space RBHF results
\cite{Wang2021,Wang2022,Tong2022ApJ}.  The
single-particle potential of neutrons $n$ and protons $p$ is
decomposed as
\begin{equation}
U_\tau(p)=U_{S,\tau}(p)+\gamma^0U_{0,\tau}(p)
+\gamma\!\cdot\!\hat p\,U_{V,\tau}(p).
\label{eq:single_particle_potential_intro}
\end{equation}
Here $\hat{\boldsymbol p}=\boldsymbol p/|\boldsymbol p|$
is the unit vector along the nucleon three-momentum in the
matter rest frame.
The full-Dirac-space calculation supplies all three Lorentz components
and avoids the particular reconstruction ambiguity associated with
positive-energy-only projections.  A local Hartree functional has no explicit
$\gamma\!\cdot\!\hat p$ term.  A central point of this work is
therefore to use the microscopic $U_V$ information explicitly before
performing the Hartree reduction, rather than discarding it.

The self-energy data are available at the density grid
$\rho=0.16,0.24,0.32,0.40$, and $0.48~{\rm fm}^{-3}$ as functions of
the isospin asymmetry, whereas the RBHF energy-per-particle data extend
down to about $0.08~{\rm fm}^{-3}$.  We use the 199 energy points
inside $0.08\leq\rho\leq0.48~{\rm fm}^{-3}$. Thirteen additional
energy points above $0.48~{\rm fm}^{-3}$, extending to
$0.50054~{\rm fm}^{-3}$, are used only as near-boundary
extrapolation diagnostics because no meson-resolved self-energy
input is available there.  Below $0.16~{\rm fm}^{-3}$, the coupling shapes
are constrained by the energy data and the smooth parametrization,
not by meson-resolved self-energies.

\section{Density-dependent Hartree framework}
\label{sec:ddrmf}

We consider nucleons coupled to effective
isoscalar-scalar, isoscalar-vector, isovector-scalar, and
isovector-vector meson fields.  The interaction Lagrangian is
\bea
{\cal L}_{\rm int}
&=&-g_\sigma(\rho)\bar\psi\sigma\psi
-g_\omega(\rho)\bar\psi\gamma_\mu\omega^\mu\psi
\nonumber\\
&&-g_\rho(\rho)\bar\psi\gamma_\mu\vec\rho^{\,\mu}\!\cdot\!\vec\tau\psi
-g_\delta(\rho)\bar\psi\vec\delta\!\cdot\!\vec\tau\psi ,
\label{eq:lint}
\eea
where the total baryon density is $\rho=\rho_n+\rho_p$, with $\rho_n$ and
$\rho_p$ the neutron and proton densities, arrows denote vectors in isospin
space, and $\vec\tau$ are the isospin Pauli matrices.
With $C_i(\rho)$ defined by Eq.~\eqref{eq:cheb_intro}, the direct
Hartree self-energies in uniform matter are
\bea
\Sigma_{S,n}&=&-C_\sigma\rho_s-C_\delta\rho_{s3},\quad
\Sigma_{S,p}=-C_\sigma\rho_s+C_\delta\rho_{s3},
\label{eq:scalar_hartree}\\
\Sigma_{0,n}^{(0)}&=&C_\omega\rho+C_\rho\rho_3,\quad
\Sigma_{0,p}^{(0)}=C_\omega\rho-C_\rho\rho_3 ,
\label{eq:vector_hartree}
\eea
where
$\rho_s=\rho_{s,n}+\rho_{s,p}$,
$\rho_{s3}=\rho_{s,n}-\rho_{s,p}$, and
$\rho_3=\rho_n-\rho_p$.
Introducing the isospin asymmetry $ \alpha=(\rho_n-\rho_p)/{\rho}, $
the neutron/proton densities are given by
$\rho_{n,p}=[(1\pm\alpha)/2]\rho$ and the respective Fermi momenta are
$k_{F,\tau}=(3\pi^2\rho_\tau)^{1/3}$.  The scalar density is
\begin{equation}
\rho_{s,\tau}=\frac{1}{\pi^2}\int_0^{k_{F,\tau}}\!dk\,
\frac{k^2 M_\tau^*}{\sqrt{k^2+M_\tau^{*2}}},
\label{eq:scalar_density}
\end{equation}
where $M_\tau^*=M+\Sigma_{S,\tau}$ and $M$ is the bare nucleon mass.
Because the vertices depend on the total baryon density, the common
rearrangement self-energy is
\bea
\Sigma_R
&=&\frac12 C_\omega' \rho^2
+\frac12 C_\rho' \rho_3^2
-\frac12 C_\sigma' \rho_s^2
-\frac12 C_\delta' \rho_{s3}^2 .
\label{eq:rearrangement}
\eea
Here a prime denotes differentiation with respect to $\rho$.
The expressions in Eq.~\eqref{eq:vector_hartree} are the \emph{direct}
vector fields only. The full time-like vector self-energies of the
density-dependent Hartree theory are
$\Sigma_{0,n}=\Sigma_{0,n}^{(0)}+\Sigma_R$ and
$\Sigma_{0,p}=\Sigma_{0,p}^{(0)}+\Sigma_R$.
Thus rearrangement is part of the vector self-energies and enters
both the single-particle chemical potentials and the pressure through
$\rho\Sigma_R$, although it does not appear explicitly in the energy
density. Its inclusion is required for thermodynamic consistency.

\section{Reduction of the full-Dirac-space self-energies}
\label{sec:mapping}

At the Fermi surface the microscopic RBHF dispersion relation can be
written as
\bea
E_\tau
=U_{0,\tau}
+\sqrt{[k_{F,\tau}+U_{V,\tau}]^2+[M+U_{S,\tau}]^2}.
\label{eq:rbhf_dispersion}
\eea
Introducing
$
\eta_\tau=U_{V,\tau}(k_{F,\tau})/k_{F,\tau},
$
the reduced self-energies at this momentum can be written as
\bea
U^{\rm red}_{S,\tau}
=
\frac{U_{S,\tau}-M\eta_\tau}{1+\eta_\tau},\quad
U^{\rm red}_{0,\tau}
=
\frac{U_{0,\tau}+E_\tau\eta_\tau}{1+\eta_\tau}.
\label{eq:Ured}
\eea
Eq.~\eqref{eq:Ured} is an exact rewriting at the Fermi surface
and identifies the quantities to be compared with local Hartree fields.
It does not eliminate the momentum dependence of the RBHF self-energies
or define a unique mapping of the microscopic energy functional onto
a local Hartree functional.

The four Hartree coupling combinations are then extracted as
\bea
C_\sigma(\rho,\alpha)
=-\frac{U^{\rm red}_{S,n}+U^{\rm red}_{S,p}}{2\rho_s},
\;
C_\omega(\rho,\alpha)
=
\frac{U^{\rm red}_{0,n}+U^{\rm red}_{0,p}}{2\rho},
\label{eq:coupling_isoscalar}
\\
C_\delta(\rho,\alpha) 
=-
\frac{U^{\rm red}_{S,n}-U^{\rm red}_{S,p}}{2\rho_{s3}},
\;
C_\rho(\rho,\alpha)
=
\frac{U^{\rm red}_{0,n}-U^{\rm red}_{0,p}}{2\rho_3}.
\label{eq:coupling_isovector}
\eea
The microscopic RBHF self-energies are matched here to the
\emph{direct} Hartree self-energies.  We exclude
$\lvert\alpha\rvert<0.05$ from the pointwise isovector extraction,
because $\rho_3, \rho_{s 3} \rightarrow 0$.  The rearrangement
contribution in Eq.~\eqref{eq:rearrangement} is excluded from
the direct-field matching in Eqs.~\eqref{eq:coupling_isoscalar}
and \eqref{eq:coupling_isovector}. In particular, the microscopic
$U^{\rm red}_{0,\tau}$ is identified with
$\Sigma^{(0)}_{0,\tau}$, not with the full
$\Sigma_{0,\tau}=\Sigma^{(0)}_{0,\tau}+\Sigma_R$.
This convention follows the separation of direct and
rearrangement fields in density-dependent Hartree theory
\cite{Hofmann2001}, and does not imply equality of the full
quasiparticle energies: the final
self-consistent rearrangement field reaches $-25.6$~MeV at
$\rho=0.40~{\rm fm}^{-3}$ for $\alpha=0.4$.

At each density we use the median over the available asymmetry points
as the density-only microscopic baseline.  The central $68\%$ interval
over $\alpha$ is used as a diagnostic of the information lost in
reducing the momentum- and asymmetry-dependent RBHF self-energy to
four density-only Hartree channels.

The role of $U_V$ is numerically important.  If it is omitted, the
density-only least-squares residual of the scalar self-energy rises
from about $1$ MeV at $\rho=0.16~{\rm fm}^{-3}$ to about $84$ MeV at
$0.48~{\rm fm}^{-3}$.  After the reduction \eqref{eq:Ured}, the
corresponding residual (RMS over neutron/proton scalar
self-energies over the sampled asymmetries at fixed density)
remains below $4$ MeV at all five densities. This estimate refers
\emph{only} to the microscopic projection, before thermodynamic
calibration: final CE-ME1 scalar-field residuals can reach about
$20$~MeV. The strong high-density growth
of the directly projected scalar coupling is largely an artifact of
discarding the space-like vector component.

\section{Chebyshev representation and microscopic baseline}
\label{sec:chebyshev}

We map the calibrated density interval onto $[-1,1]$ according to
\bea
z(\rho)=
\frac{2\rho-(\rho_{\rm min}+\rho_{\rm max})}
{\rho_{\rm max}-\rho_{\rm min}},
\label{eq:z_map}
\eea
where $\rho_{\rm min}=0.08~{\rm fm}^{-3}$ and
$\rho_{\rm max}=0.48~{\rm fm}^{-3}$.
The final functional uses the cubic representation
\bea
C_i(\rho)=a_{i0}+a_{i1}T_1(z)+a_{i2}T_2(z)+a_{i3}T_3(z),
\label{eq:cubic_cheb}
\eea
with
$T_1=z$, $T_2=2z^2-1$, and $T_3=4z^3-3z$.
The continuous Chebyshev orthogonality relation,
\bea
\int_{-1}^{1}\frac{T_m(z)T_n(z)}{\sqrt{1-z^2}}\,dz=0,
\qquad m\ne n ,
\label{eq:orthogonality}
\eea
is used to organize the spectral expansion, but it does not make
the discrete, nonlinear RBHF fit orthogonal: correlations among fitted
coefficients depend on the data, imposed constraints, and the
nonlinear form of the functional itself.  We use the Chebyshev basis for its boundedness, hierarchical
form, and analytic derivatives, not as an argument for statistical
independence.  The $U_V$-reduced microscopic baseline is listed in
Table~\ref{tab:UVcouplings}.  The couplings are smooth: $C_\sigma$ and
$C_\omega$ decrease with density, $C_\delta$ increases moderately, and
$C_\rho$ decreases.  The asymmetry-induced relative half-width is
below $1.2\%$ for the isoscalar channels, while it reaches about
$18\%$ for $C_\delta$ at the lowest density and remains at the
few-percent level over most of the interval.

\begin{table}[t]
\centering
\small
\caption{$U_V$-reduced microscopic coupling combinations
$C_i=(g_i/m_i)^2$ obtained from the full-Dirac-space RBHF
self-energies.  All entries are in ${\rm GeV}^{-2}$.}
\label{tab:UVcouplings}
\begin{tabular}{c|rrrr}
\hline
$\rho\,[{\rm fm}^{-3}]$ & $C_\sigma$ & $C_\omega$ & $C_\delta$ & $C_\rho$\\
\hline
0.16 & 298.78 & 210.51 & 15.56 & 30.80\\
0.24 & 280.56 & 199.75 & 19.67 & 24.64\\
0.32 & 268.40 & 186.99 & 22.17 & 22.07\\
0.40 & 261.92 & 174.76 & 25.05 & 20.28\\
0.48 & 252.65 & 162.42 & 25.85 & 18.72\\
\hline
\end{tabular}
\end{table}

\section{\texorpdfstring{Calibration to RBHF energies and saturation properties}{Calibration to RBHF energies and saturation properties}}
\label{sec:thermorefit}

The reduced self-energy projection provides the meson-re\-solved
microscopic baseline, but matching self-energies at the Fermi surface
is not identical to matching the RBHF energy functional.  We therefore
apply a second, intentionally low-dimensional calibration to the
RBHF energies and the saturation properties of cold nuclear matter.  The cubic functions are required to remain positive and
monotonic in the same sense as the microscopic input over the full
interval. These restrictions suppress unphysical oscillations and
limit, without eliminating, compensating adjustments of the scalar
and vector channels.

The 16 cubic coefficients are fitted sequentially, with each of
the 199 energy observations carrying equal weight. Five equality constraints, together with positivity and monotonicity
conditions, restrict the admissible parameter space; they do not
make the remaining coefficient combinations statistically independent. We first calibrate
$C_\sigma$ and $C_\omega$ to the
symmetric-matter energy per particle $E/A$, imposing the microscopic
saturation parameters
\bea
\rho_{\rm sat}=0.188~{\rm fm}^{-3},\quad
E_{\rm sat}=-15.40~{\rm MeV},
\label{eq:hard_saturation}
\eea
together with $P(\rho_{\rm sat})=0$ and incompressibility
$K_\infty=258$ MeV.  Second, these isoscalar functions are held fixed and $C_\delta$
and $C_\rho$ are determined from the five asymmetric data sets.
We define the quadratic symmetry energy by
\begin{equation}
\frac{E}{A}(\rho,\alpha)=\frac{E}{A}(\rho,0)
 +S_2(\rho)\alpha^2+\mathcal O(\alpha^4),
\end{equation}
where $S_2(\rho)=\tfrac12\left.\partial^2(E/A)/\partial\alpha^2
\right|_{\alpha=0}$. Its saturation values
$J=S_2(\rho_{\rm sat})$ and
$L=3\rho_{\rm sat}S_2'(\rho_{\rm sat})$ are fixed to
\bea J=33.1~{\rm MeV},\qquad L=65.2~{\rm
  MeV}.
\label{eq:JLtargets}
\eea
These values are taken from the published Bonn-A full-Dirac-space RBHF
calculation \cite{Tong2022ApJ}.  In symmetric matter the quadratic coefficient can be decomposed
into a kinetic term and the $\rho$- and $\delta$-meson responses:
\begin{equation}
S_2(\rho)=\frac{k_F^2}{6E_F^*}+\frac12 C_\rho\rho
-\frac{C_\delta\rho}{2}\,
\frac{(M^*/E_F^*)^2}{1+C_\delta A},
\label{eq:symmetry_analytic}
\end{equation}
where $E_F^*=\sqrt{k_F^2+M^{*2}}$,
$k_F=(3\pi^2\rho/2)^{1/3}$, and
$A=(2/\pi^2)\int_0^{k_F} dk\,k^4/(k^2+M^{*2})^{3/2}$ (in
natural units).
The curvature
$K_{\rm sym}=9\rho_{\rm sat}^2S_2''(\rho_{\rm sat})$ is not used in
calibration.  The energy residuals and proximity to the smooth
microscopic baseline select the remaining isovector freedom under the
positivity and monotonicity constraints. 
\begin{table}[t]
\centering
\small
\caption{Final CE-ME1 coefficients in
$C_i(\rho)=\sum_{n=0}^{3}a_{in}T_n[z(\rho)]$.
All entries are in ${\rm GeV}^{-2}$.}
\begin{tabular}{c|rrrr}
\hline
$i$ & $a_{i0}$ & $a_{i1}$ & $a_{i2}$ & $a_{i3}$\\
\hline
$\sigma$ & 272.600466 & -38.723320 &  2.119386 &  0.745273\\
$\omega$ & 195.097250 & -30.905766 &  0.173863 &  2.168564\\
$\delta$ &  20.602444 &   7.137289 & -1.594963 & -0.079499\\
$\rho$   &  30.501194 &  -8.418124 &  3.878273 & -0.862753\\
\hline
\end{tabular}
\label{tab:final-CE-ME1}
\end{table}

The final CE-ME1 coefficients are given in Table~\ref{tab:final-CE-ME1}.
They reproduce the fitted saturation quantities
$E_{\rm sat}$, $K_\infty$, $J$, and $L$, and predict
$K_{\rm sym}=33.14$ MeV. The combined RMS deviation
from the $N=199$ RBHF energy points is $0.6127$ MeV,
with equal weight assigned to each point.
Since the RBHF data carry no pointwise uncertainties,
this RMS measures the discrepancy between the functional
and the microscopic input rather than a statistical
goodness of fit. The six asymmetry sets contain
$N_\alpha=(36,39,30,35,30,29)$ points, with individual
RMS deviations increasing from $0.413$ MeV at
$\alpha=0$ to $0.745$ MeV at $\alpha=1$.
For the sampled fixed-asymmetry sequences above
saturation, $dP/d\rho$ remains positive and
$c_s^2\leq 0.797$, indicating positive compressibility
and subluminal sound speeds throughout the calibrated
density interval.

\begin{figure}[hbt]
\centering
\includegraphics[width=\columnwidth]{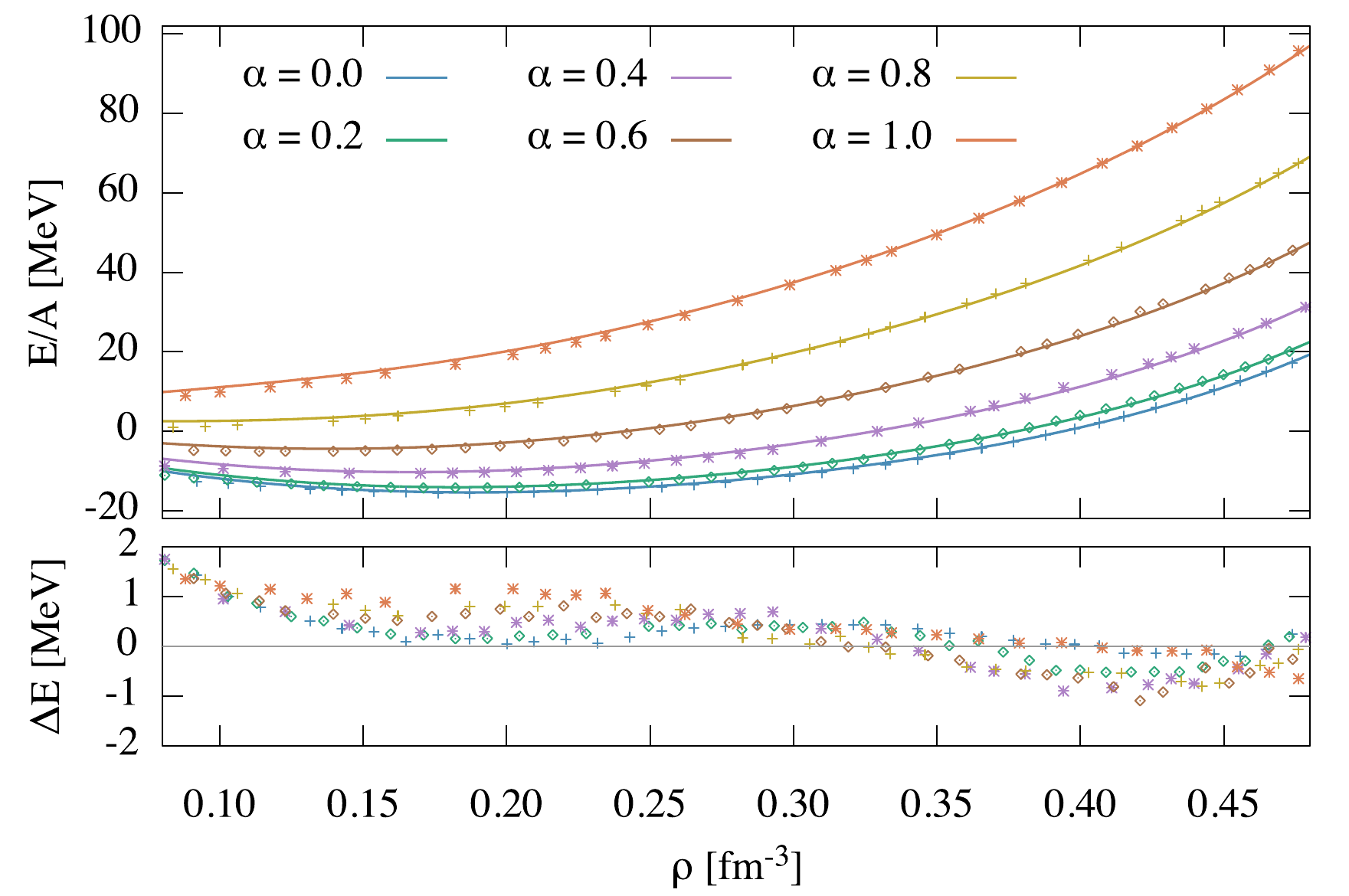}
\caption{RBHF energy per particle and the final smooth CE-ME1
calibration for $\alpha=0,0.2,\ldots,1$ (upper panel), and residuals
$(E/A)_{\rm CE-ME1}-(E/A)_{\rm RBHF}$ (lower panel).  Only points in
$0.08\leq\rho\leq0.48~{\rm fm}^{-3}$ enter the quoted RMS.}
\label{fig:rbhf_fit}
\end{figure}
Fig.~\ref{fig:rbhf_fit} shows the RBHF energies and the final CE-ME1
curves.  The residuals have a systematic density dependence rather than noise-like
scatter; the largest fitted absolute deviation is about $1.76$ MeV.

\begin{figure}[hbt]
\centering
\includegraphics[width=\columnwidth]{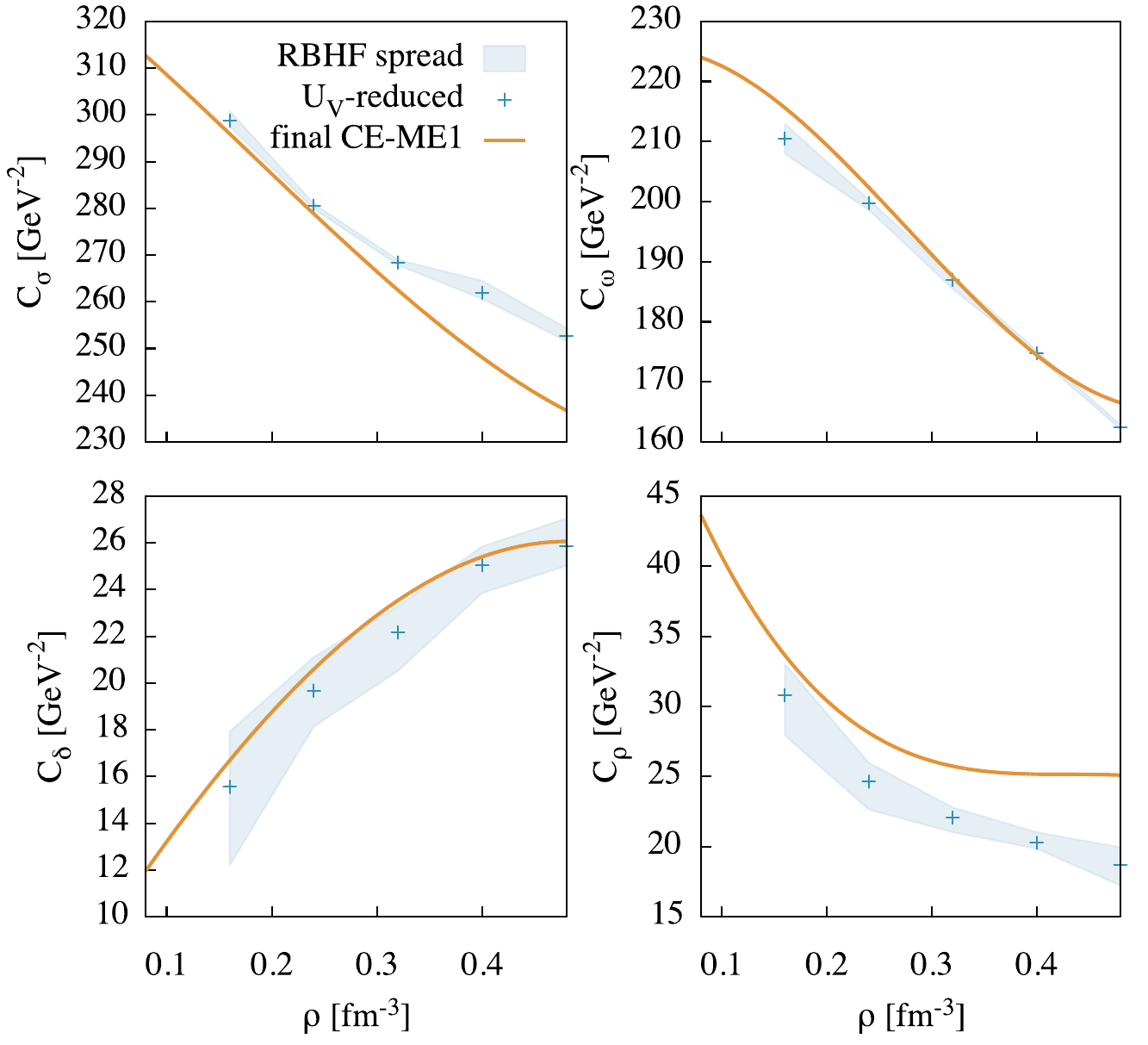}
\caption{Final CE-ME1 coupling functions (solid curves) compared with
the $U_V$-reduced full-Dirac-space RBHF projection (points).  Shaded
bands show the central $68\%$ variation with asymmetry of the
digitized microscopic mapping (not statistical error bands).}
\label{fig:coupling_comparison}
\end{figure}

Fig.~\ref{fig:coupling_comparison} compares the final functions with
the $U_V$-reduced microscopic projection.  At the five microscopic densities the final $C_\sigma$ and
$C_\omega$ differ from the baseline by up to $6.3\%$ and
$2.6\%$, respectively; for the scalar field this corresponds to
differences of order $20$~MeV at the highest density.  The final $C_\delta$ remains within about
$7.3\%$ of the microscopic values.  The largest adjustment occurs in $C_\rho$, reaching about $34\%$
at $0.48~{\rm fm}^{-3}$. The near-plateau of $C_\rho$ is consistent with the imposed
monotonicity requirement. Without a relaxed refit, however, its
quantitative origin cannot be attributed to that constraint alone.

\subsection{Beta-equilibrated matter and comparison with DD-ME2}
\label{subsec:beta_comparison}

\begin{figure}[!tbp]
\centering
\includegraphics[width=\columnwidth]{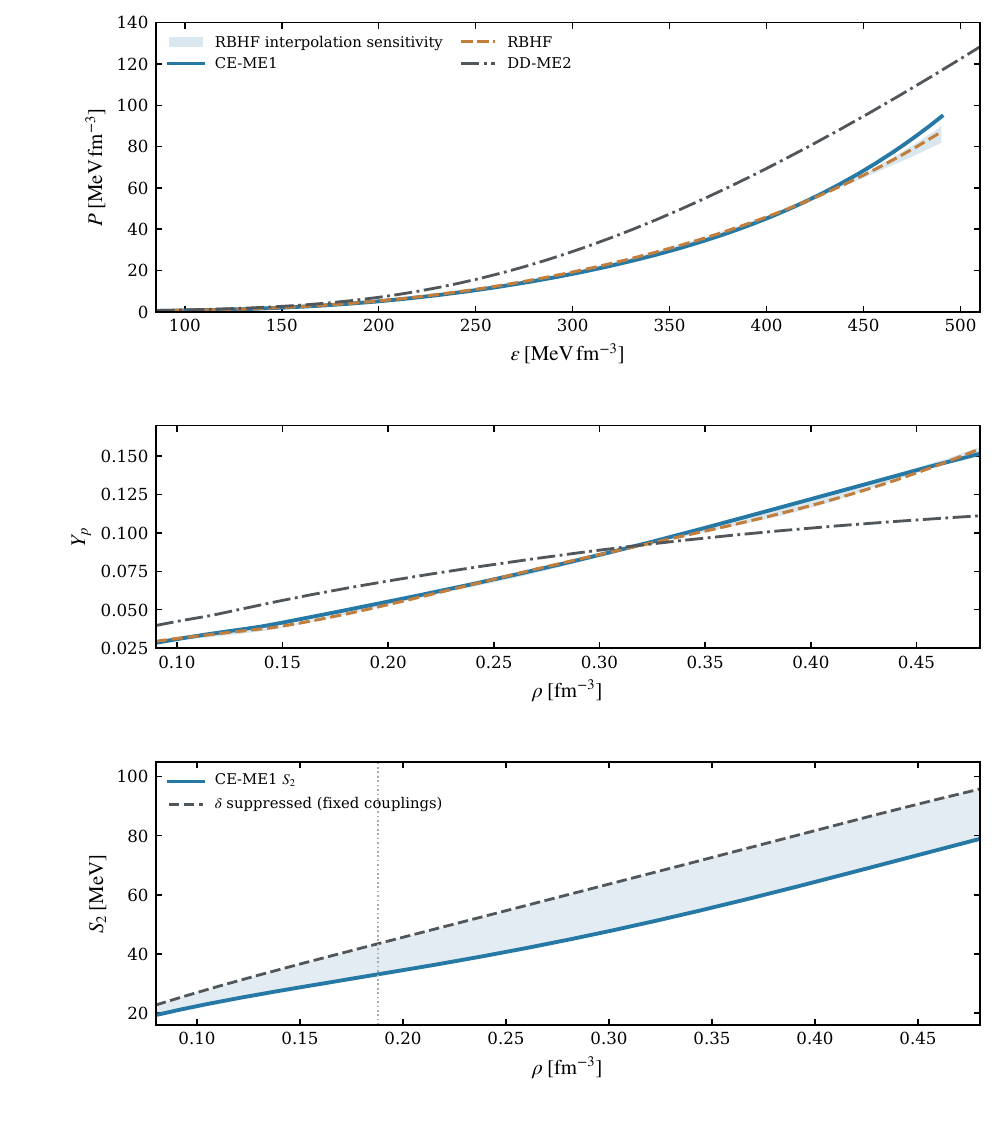}
\caption{Cold charge-neutral beta-equilibrated matter
and quadratic symmetry energy. Upper panel:
pressure as a function of total energy density for CE-ME1, the smooth
interpolation of the full-Dirac-space RBHF $E/A$ data, and DD-ME2. The
shaded band shows the sensitivity of the RBHF pressure to nearby smooth
representations used before differentiation. Middle panel:
corresponding proton fractions.
Lower panel: $S_2(\rho)$ of CE-ME1 (solid) and the
fixed-coupling diagnostic with the $\delta$ contribution suppressed
(dashed). The shaded separation shows the negative $\delta$ contribution;
the dashed curve is not a recalibrated $\delta$-free model.
The beta-equilibrium comparison is restricted to the CE-ME1 calibration
range and is not extrapolated to neutron-star central densities.}
\label{fig:beta_comparison}
\end{figure}

We next test the parametrization in cold charge-neutral $npe\mu$
matter, focusing on its equation of state and proton fraction,
see Fig.~\ref{fig:beta_comparison}. The same figure also shows
the quadratic symmetry energy of the calibrated functional and
exhibits the impact of the isovector-scalar $\delta$ channel.  As a phenomenological benchmark we
include the same results obtained using the DD-ME2
parametrization~\cite{Lalazissis2005}.  Our implementation of the DD-ME2
model reproduces its standard saturation properties,
$\rho_0=0.152~{\rm fm}^{-3}$, $E/A=-16.14$ MeV, $K\simeq251$ MeV,
$J\simeq32.3$ MeV, and $L\simeq51.3$ MeV~\cite{Sedrakian2023}.

For CE-ME1 the
proton fraction increases from $Y_p=0.052$ at the RBHF saturation
density to $0.152$ at $0.48~{\rm fm}^{-3}$.  Along this sequence we
find
$\min(dP/d\rho)=64.0~\mathrm{MeV},\qquad \max c_s^2=0.753,$
where $c_s$ is the speed of sound. The baryonic curvature matrix
$\partial\mu_i/\partial\rho_j$ is also positive definite on a grid
spanning $0\leq\alpha\leq0.9$ above saturation, with a minimum
eigenvalue of approximately $305~{\rm MeV\,fm^3}$.  We do not impose
mechanical stability below saturation, where a liquid--gas spinodal is
physical, and no crust construction is attempted here.

For a direct microscopic comparison we also construct
beta-equilibrated matter from the same RBHF energy-per-particle data
used in the calibration.  To ensure an accurate density derivative
needed for the pressure, we represent the 199 RBHF energies in the
calibration domain by a smooth tensor-product surface, using Chebyshev
polynomials through order five in density and a cubic polynomial in
$\alpha^2$.  The representation has an RMS deviation of $0.097$ MeV
from the original RBHF energies.  The sensitivity band in
Fig.~\ref{fig:beta_comparison} is obtained by varying the density
order from four to seven and the $\alpha^2$ order from two to four.
This band should be read as an interpolation/differentiation
sensitivity, not as a statistical uncertainty.

Fig.~\ref{fig:beta_comparison} shows the pressure and proton
fraction for CE-ME1, DD-ME2, and the reconstructed RBHF EoS.  The pressure agrees
reasonably up to about $0.4~{\rm fm}^{-3}$: at that density the pressures are $44.4$ and
$45.3~{\rm MeV\,fm}^{-3}$, respectively.  At the upper edge of the
calibrated interval CE-ME1 gives $P=94.6~{\rm MeV\,fm}^{-3}$, compared
with the central smooth RBHF value $87.3~{\rm MeV\,fm}^{-3}$; the
latter varies between approximately $82$ and $90~{\rm MeV\,fm}^{-3}$
under the interpolation changes described above. The CE-ME1 edge
pressure is about $8.4\%$ above the central reconstructed RBHF
value and lies outside this interpolation-sensitivity band. The corresponding
proton fractions are $0.152$ and $0.155$.  DD-ME2 is appreciably
stiffer over the same interval, reaching $P=128.1~{\rm MeV\,fm}^{-3}$
at $0.48~{\rm fm}^{-3}$, and has a smaller high-density proton
fraction, $Y_p=0.111$.

The full density dependence $S_2(\rho)$ is shown in the lower
panel of Fig.~\ref{fig:beta_comparison}, alongside a case
in which the $\delta$ contribution is suppressed while all other
couplings are held fixed. At $\alpha=0$, the $\delta$ mean field
vanishes, but its isovector response contributes negatively to
$S_2$: $-10.38$~MeV at $\rho_{\rm sat}$ and approximately
$-16.86$~MeV at $0.48~{\rm fm}^{-3}$.
(Note that the illustrative result shown without the $\delta$ meson does not result from  a
 refitted functional and is not intended as
a realistic alternative description of nuclear matter.)

\section{Discussion}
\label{sec:discussion}

The main improvement of the CE-ME1 construction over a direct
$U_S$--$U_0$ projection is the explicit use of the
space-like RBHF self-energy.  At $\rho=0.48~{\rm fm}^{-3}$ its
magnitude is comparable with a sizable fraction of the Fermi momentum,
so omitting it strongly distorts the scalar field inferred for a
momentum-independent Hartree theory.  Once $U_V$ is absorbed at the
Fermi surface, all four microscopic coupling combinations become
smooth and can be represented at cubic order.

A residual difference between the Fermi-surface self-energy mapping
and the RBHF energy per particle is nevertheless expected.  The full
RBHF self-energies retain momentum dependence over the occupied Fermi
sea, while the local Hartree functional is momentum independent.
Consequently, matching the Lorentz components at one momentum is not
an energy-functional identity.  The thermodynamic calibration should
therefore be interpreted as a controlled correction to the microscopic
local projection, not as a new independent determination of the meson
channels.

The cubic Chebyshev representation spans the ordinary cubic-polynomial
space but offers bounded basis functions and analytic rearrangement
derivatives. It does not remove correlations among the nonlinear fit
coefficients. Higher-order constrained fits and convergence tests remain
for future work; the reported asymmetry spreads and residuals are
model-reduction diagnostics only. 

The beta-equilibrium comparison (Fig.~\ref{fig:beta_comparison})
checks derivative and composition consistency of the thermodynamic
reduction using the same microscopic energies.  The increasing
pressure difference near $0.48~{\rm fm}^{-3}$ is consistent with the
larger derivative sensitivity of the energy fit. The 13 unused
near-boundary points, $0.48<\rho\leq0.50054~{\rm fm}^{-3}$, have an
RMS residual of $0.837$~MeV and a maximum absolute residual of
$1.710$~MeV.  The relatively high saturation density
$\rho_{\rm sat}=0.188~{\rm fm}^{-3}$ is inherited from the
Bonn-A RBHF calculation; no phenomenological saturation density
is imposed.  The DD-ME2 CDF, as a functional widely used in
astrophysical applications, provides a phenomenological reference with
a different saturation point and a markedly stiffer high-density
equation of state.

The calibration is restricted to the density range
$0.08\leq\rho\leq0.48~{\rm fm}^{-3}$. Extrapolating to nuclear
surfaces, neutron-star crusts, or the central densities of massive
stars would require additional microscopic input at low or high
density. Several effects could improve the calibration of the
functional: many-body contributions beyond the ladder summation,
inelastic scattering at high energies, and short-range correlations.
Momentum-dependent or Hartree--Fock extensions of the CDF developed
here would also help, since they retain more of the microscopic input.
Another route to a better functional is optimization against
finite-nucleus data, which we leave for future work.

\section{Conclusion}
\label{sec:conclusions}

We have constructed the CE-ME1 density-dependent covariant functional
from full-Dirac-space RBHF input.  The microscopic $U_S$, $U_0$, and
$U_V$ components are first reduced at the Fermi surface to effective
scalar and time-like self-energies and then mapped onto the four-meson
$\sigma$-$\omega$-$\delta$-$\rho$ Hartree channels.  Including $U_V$
is essential: it removes the strong high-density growth of the scalar
coupling found when the space-like component is neglected, and it
reduces the residual of the local self-energy mapping to a few
MeV. The density dependence of the couplings is represented by cubic
Chebyshev polynomials on $0.08\leq\rho\leq0.48~{\rm fm}^{-3}$.  A
sequential isoscalar/isovector calibration to 199 RBHF
energy-per-particle values reproduces the microscopic saturation point
exactly and gives $K_\infty=258.0$ MeV, $J=33.10$ MeV, and $L=65.20$
MeV, with a combined energy RMS of $0.613$ MeV. The couplings remain
positive and monotonic throughout the interval, and uniform matter is
mechanically stable and causal above saturation.  In beta equilibrium
the CE-ME1 pressure and proton fraction agree reasonably with the
same-data RBHF interpolation over most of the range, while the
pressure discrepancy grows near $0.48~{\rm fm}^{-3}$.  Over the common comparison interval, DD-ME2 predicts a
substantially stiffer equation of state than CE-ME1.

The curvature $K_{\rm sym}=9\rho_{\rm sat}^2S_2''(\rho_{\rm sat})$ is
not used in the calibration. CE-ME1 predicts $33.14$~MeV, whereas the
\emph{published} full-Dirac-space Bonn-A RBHF value is $-67.3$~MeV
\cite{Tong2022ApJ}. The value cannot reliably be recovered from second
derivatives of our data interpolation, which are strongly order
dependent. The discrepancy of $\simeq100$~MeV illustrates how sensitively
this second derivative probes the low-order local representation; the
restricted isovector shapes and the discarded momentum dependence may
both play roles, without a separate quantitative attribution.  We
therefore regard $K_{\rm sym}$ as a sensitivity test of the
reduction. 

The present construction shows that a low-order Chebyshev representation
can retain much of the microscopic Lorentz structure while providing a
compact, thermodynamically consistent CDF. Additional full-Dirac-space
self-energy points, especially above $0.48~{\rm fm}^{-3}$, will be needed
before CE-ME1 can be extended to high-density neutron-star applications.
Such calculations may, however, fail to converge or to yield reliable
results at these densities. In that case, one would have to resort to
extrapolation schemes based on polytropes or speed-of-sound
parametrizations.

In closing, it is instructive to compare the construction with that of
Alford et al.~\cite{Alford2022}, who also calibrate a relativistic
mean-field functional using microscopic uniform-matter input.  The two
approaches differ in their microscopic input, functional form, and
density range.  Ref.~\cite{Alford2022} fits a density-independent
$\sigma$-$\omega$-$\rho$ Lagrangian with nonlinear meson interactions
to chiral-EFT pure neutron matter over $0.5\leq n/n_0\leq 2$,
supplemented by empirical saturation properties. Its extension to
higher densities is determined by the adopted functional form and
constrained by astrophysical observations. CE-ME1 instead employs
density-dependent $\sigma$-$\omega$-$\delta$-$\rho$ couplings derived
from full-Dirac-space RBHF self-energies, including $U_V$, and
energies at six isospin asymmetries. Its microscopic calibration
extends to $\rho=0.48~\mathrm{fm}^{-3}$, corresponding to about $2.55$
times the RBHF saturation density, but no extrapolation beyond this
range is attempted. The two approaches are therefore complementary:
CE-ME1 is anchored in the Lorentz-resolved self-energies of
relativistic RBHF theory~\cite{Wang2021,Wang2022,Tong2022ApJ}, whereas
the functional of Ref.~\cite{Alford2022} is constrained by
nonrelativistic chiral-EFT calculations of neutron matter. The latter
is extended beyond the microscopic density range within its
relativistic mean-field framework, subject to additional astrophysical
constraints. A comparable extension of CE-ME1 to higher densities will
require confronting it with neutron-star observations, which we leave
for future work.

\nolinenumbers
\section*{Acknowledgments}

This work was funded by Deutsche Forschungsgemeinschaft Grant
No.~SE~1836/6-1 and Polish NCN Grant No.~2023/51/B/ST9/02798.

\section*{Declaration of competing interest}
The author declares that there are no known competing financial
interests or personal relationships that could have appeared to
influence the work reported in this paper.

\section*{Data availability}
Data will be made available on request.

\setlength{\bibsep}{0pt}

\begin{thebibliography}{10}
\expandafter\ifx\csname url\endcsname\relax
  \def\url#1{\texttt{#1}}\fi
\expandafter\ifx\csname urlprefix\endcsname\relax\def\urlprefix{URL }\fi
\expandafter\ifx\csname href\endcsname\relax
\def\href#1#2{#2} \def\path#1{#1}\fi


\bibitem{Serot1997}
B.~D.~Serot and J.~D.~Walecka,
Recent progress in quantum hadrodynamics,
Int. J. Mod. Phys. E 6, (1997)  515--631.
\newblock \href {https://doi.org/doi:10.1142/S0218301397000299}
  {\path{doi:10.1142/S0218301397000299}}.

\bibitem{Ring1996}
P.~Ring, Relativistic mean field theory in finite nuclei, Progress in Particle
  and Nuclear Physics 37 (1996) 193--263.
\newblock \href {https://doi.org/10.1016/0146-6410(96)00054-3}
  {\path{doi:10.1016/0146-6410(96)00054-3}}.

\bibitem{Meng2006}
J.~Meng, H.~Toki, S.~Zhou, S.~Zhang, W.~Long, L.~Geng, Relativistic continuum
  Hartree Bogoliubov theory for ground-state properties of exotic nuclei,
  Progress in Particle and Nuclear Physics 57~(2) (2006) 470--563.
\newblock \href {https://doi.org/10.1016/j.ppnp.2005.06.001}
  {\path{doi:10.1016/j.ppnp.2005.06.001}}.

\bibitem{Sedrakian2023}
A.~{Sedrakian}, J.~J. {Li}, F.~{Weber}, {Heavy baryons in compact stars},
  Progress in Particle and Nuclear Physics 131 (2023) 104041.
\newblock \href {http://arxiv.org/abs/2212.01086} {\path{arXiv:2212.01086}},
  \href {https://doi.org/10.1016/j.ppnp.2023.104041}
  {\path{doi:10.1016/j.ppnp.2023.104041}}.

\bibitem{Typel1999}
S.~Typel, H.~Wolter, Relativistic mean field calculations with
  density-dependent meson-nucleon coupling, Nuclear Physics A 656~(3-4) (1999)
  331--364.
\newblock \href {https://doi.org/10.1016/s0375-9474(99)00310-3}
  {\path{doi:10.1016/s0375-9474(99)00310-3}}.

\bibitem{Dalen2005}
E.~N. E.~v. Dalen, C.~Fuchs, A.~Faessler, Momentum, density, and isospin
  dependence of symmetric and asymmetric nuclear matter properties, Physical
  Review C 72~(6) (2005) 065803.
\newblock \href {https://doi.org/10.1103/physrevc.72.065803}
  {\path{doi:10.1103/physrevc.72.065803}}.

\bibitem{Gogelein2008}
P.~G\"ogelein, E.~N.~E.~v. Dalen, C.~Fuchs, H.~M\"uther, Nuclear matter in the
  crust of neutron stars derived from realistic {NN} interactions, Physical
  Review C 77~(2) (2008) 025802.
\newblock \href {https://doi.org/10.1103/physrevc.77.025802}
  {\path{doi:10.1103/physrevc.77.025802}}.

\bibitem{Giai2010}
N.~Van~Giai, B.~V. Carlson, Z.~Ma, H.~Wolter, The
 Dirac–Brueckner–Hartree–Fock approach: from infinite matter to
  effective lagrangians for finite systems, Journal of Physics G: Nuclear and
  Particle Physics 37~(6) (2010) 064043.
\newblock \href {https://doi.org/10.1088/0954-3899/37/6/064043}
  {\path{doi:10.1088/0954-3899/37/6/064043}}.

\bibitem{Typel2018}
S.~{Typel}, {Relativistic Mean-Field Models with Different Parametrizations of
  Density Dependent Couplings}, Particles 1~(1) (2018) 2.
\newblock \href {https://doi.org/10.3390/particles1010002}
  {\path{doi:10.3390/particles1010002}}.

\bibitem{Wang2021}
S.~Wang, Q.~Zhao, P.~Ring, J.~Meng, Nuclear matter in relativistic
  Brueckner-Hartree-Fock theory with Bonn potential in the full Dirac space,
  Physical Review C 103~(5) (2021) 054319.
\newblock \href {https://doi.org/10.1103/physrevc.103.054319}
  {\path{doi:10.1103/physrevc.103.054319}}.

\bibitem{Wang2022}
S.~Wang, H.~Tong, Q.~Zhao, C.~Wang, P.~Ring, J.~Meng, Asymmetric nuclear matter
  and neutron star properties in relativistic ab initio theory in the full
  Dirac space, Physical Review C 106~(2) (2022) L021305.
\newblock \href {https://doi.org/10.1103/PhysRevC.106.L021305}
  {\path{doi:10.1103/PhysRevC.106.L021305}}.

\bibitem{Tong2022ApJ}
H.~Tong, C.~Wang, S.~Wang, Nuclear matter and neutron stars from relativistic
  Brueckner–Hartree–Fock theory, The Astrophysical Journal 930~(2) (2022)
  137.
\newblock \href {https://doi.org/10.3847/1538-4357/ac65fc}
  {\path{doi:10.3847/1538-4357/ac65fc}}.

\bibitem{Hofmann2001}
F.~Hofmann, C.~M. Keil, H.~Lenske, Density dependent hadron field theory for
  asymmetric nuclear matter and exotic nuclei, Physical Review C 64~(3) (2001)
  034314.
\newblock \href {https://doi.org/10.1103/physrevc.64.034314}
  {\path{doi:10.1103/physrevc.64.034314}}.

\bibitem{Lalazissis2005}
G.~A. Lalazissis, T.~Nikšić, D.~Vretenar, P.~Ring, New relativistic
  mean-field interaction with density-dependent meson-nucleon couplings,
  Physical Review C 71~(2) (2005) 024312.
\newblock \href {https://doi.org/10.1103/physrevc.71.024312}
  {\path{doi:10.1103/physrevc.71.024312}}.

\bibitem{Alford2022}
M.~G. Alford, L.~Brodie, A.~Haber, I.~Tews, {Relativistic mean-field theories
  for neutron-star physics based on chiral effective field theory}, Phys. Rev.
  C 106~(5) (2022) 055804.
\newblock \href {http://arxiv.org/abs/2205.10283} {\path{arXiv:2205.10283}},
  \href {https://doi.org/10.1103/PhysRevC.106.055804}
  {\path{doi:10.1103/PhysRevC.106.055804}}.

\end{thebibliography}

\end{document}